\pdftrailerid{}          
\documentclass[twocolumn]{aastex7}

\usepackage[utf8]{inputenc}
\usepackage[T1]{fontenc}

\usepackage{newtxtext}
\usepackage[scaled=0.98, zerostyle=c]{newtxtt}  
\usepackage{newtxmath}

\usepackage{graphicx}
\usepackage{amsmath}
\usepackage{xspace}
\usepackage{booktabs}
\usepackage{subcaption}  

\usepackage{hyperref}
\hypersetup{
  bookmarksnumbered,  
  pdftitle={Lensing unions with UNIONS},
  pdfauthor={I. Cheng et al.},
  pdfcreator={},
  pdfproducer={},
  pdfdisplaydoctitle=true,
}

\graphicspath{{./img/}}

\shorttitle{Lensing unions with UNIONS}
\shortauthors{I.~Cheng et al.}
\newcommand*{\overbar}[1]{\mkern 1.5mu\overline{\mkern-2.0mu#1\mkern-1.5mu}\mkern 1.5mu}
\newcommand*{\errs}[2]{\ensuremath{\text{\raisebox{0.5ex}{\tiny$^{+#1}_{-#2}$}}}}%
\let\fdgold\fdg
\renewcommand*{\fdg}{\ensuremath{\hspace{0.25pt}\fdgold\hspace{-0.5pt}}}
\renewcommand*{\arcsec}{\ensuremath{''}\xspace}  
\newcommand*{\degree}{\ensuremath{^\circ}\xspace}  
\newcommand*{\h}{\ensuremath{\,h}\xspace}  
\newcommand*{\m}{\ensuremath{\,\text{m}}\xspace}  
\newcommand*{\km}{\ensuremath{\,\text{km}}\xspace}  
\newcommand*{\s}{\ensuremath{\,\text{s}}\xspace}  
\newcommand*{\Gyr}{\ensuremath{\,\text{Gyr}}\xspace}  
\newcommand*{\kpc}{\ensuremath{\,\text{kpc}}\xspace}  
\newcommand*{\Mpc}{\ensuremath{\,\text{Mpc}}\xspace}  
\newcommand*{\z}{\ensuremath{z}\xspace}  
\newcommand*{\Mstar}{\ensuremath{M_\star}\xspace}  
\newcommand*{\Msun}{\ensuremath{\,\mathrm{M}_\odot}\xspace}  
\newcommand*{\ravg}{\ensuremath{R_\text{avg}}\xspace}  
\newcommand*{\p}{\ensuremath{p}\xspace}  
\newcommand*{\dsigma}{\ensuremath{\Delta\Sigma}\xspace}  
\newcommand*{\rdsigma}{\ensuremath{R\Delta\Sigma}\xspace}  
\newcommand*{\lzr}{\ensuremath{\langle z \rangle}\xspace}  
\newcommand*{\mhalo}[1][]{\ensuremath{M_{\text{halo}#1}}\xspace}  
\renewcommand*{\c}{\ensuremath{c}\xspace}  
\newcommand*{\rvir}{\ensuremath{R_\text{vir}}\xspace}  
\newcommand*{\ratio}{\ensuremath{\mathcal{R}}\xspace}  

\begin{document}

\title{Unions with UNIONS:~Using galaxy-galaxy lensing to probe galaxy mergers}

\author[orcid=0000-0002-8618-7990]{Isaac Cheng}
\affiliation{Department of Physics and Astronomy, University of Waterloo, 200 University
Ave W, Waterloo, ON N2L 3G1, Canada}
\affiliation{Cahill Center for Astronomy and Astrophysics, California Institute of
Technology, MC 249-17, Pasadena, CA 91125, USA}
\email[show]{\href{mailto:icheng@astro.caltech.edu}{icheng@astro.caltech.edu} (IC)}
\correspondingauthor{Isaac Cheng}

\author[orcid=0000-0001-5148-9203]{Jack Elvin-Poole}
\affiliation{Department of Physics and Astronomy, University of Waterloo, 200 University
Ave W, Waterloo, ON N2L 3G1, Canada}
\affiliation{Waterloo Centre for Astrophysics, University of Waterloo, 200 University Ave
W, Waterloo, ON N2L 3G1, Canada}
\email[show]{\\\hspace*{0.46cm}%
\href{mailto:jack.elvin-poole@uwaterloo.ca}{jack.elvin-poole@uwaterloo.ca},
\href{mailto:jack.elvinpoole@gmail.com}{jack.elvinpoole@gmail.com} (JE-P)}

\author[orcid=0000-0002-1437-3786]{Michael J.~Hudson}
\affiliation{Department of Physics and Astronomy, University of Waterloo, 200 University
Ave W, Waterloo, ON N2L 3G1, Canada}
\affiliation{Waterloo Centre for Astrophysics, University of Waterloo, 200 University Ave
W, Waterloo, ON N2L 3G1, Canada}
\affiliation{Perimeter Institute for Theoretical Physics, 31 Caroline St. North, Waterloo,
ON N2L 2Y5, Canada}
\email[show]{\\\hspace*{0.46cm}%
\href{mike.hudson@uwaterloo.ca}{mike.hudson@uwaterloo.ca} (MJH)}

\author[orcid=0009-0006-3551-3641]{Ruxin Barr\'e}
\affiliation{School of Physics and Astronomy, University of Victoria, Victoria, BC,
Canada}
\affiliation{Department of Physics and Astronomy, University of Waterloo, 200 University
Ave W, Waterloo, ON N2L 3G1, Canada}
\email{fake_email_fix_later@fake-email.com}  

\author[orcid=0000-0002-1768-1899]{Sara L.~Ellison}
\affiliation{School of Physics and Astronomy, University of Victoria, Victoria, BC,
Canada}
\email{fake_email_fix_later@fake-email.com}  

\author[orcid=0000-0003-2794-8398]{Robert W.~Bickley}
\affiliation{School of Physics and Astronomy, University of Victoria, Victoria, BC,
Canada}
\email{fake_email_fix_later@fake-email.com}  

\author[orcid=0000-0001-5486-2747]{Thomas J.~L.~de Boer}
\affiliation{Institute for Astronomy, University of Hawaii, 2680 Woodlawn Drive, Honolulu
HI 96822}
\email{fake_email_fix_later@fake-email.com}  

\author[orcid=0000-0003-2239-7988]{S\'ebastien Fabbro}
\affiliation{National Research Council of Canada, Herzberg Astronomy \& Astrophysics
Research Centre, 5071 West Saanich Road, Victoria, BC V9E 2E7, Canada}
\email{fake_email_fix_later@fake-email.com}  

\author[orcid=0000-0002-8919-079X]{Leonardo Ferreira}
\affiliation{School of Physics and Astronomy, University of Victoria, Victoria, BC,
Canada}
\email{fake_email_fix_later@fake-email.com}  

\author[orcid=0009-0004-3655-4870]{Sacha Guerrini}
\affiliation{Universit\'e Paris Cit\'e, Universit\'e Paris-Saclay, CEA, CNRS, AIM, 91191,
Gif-sur-Yvette, France}
\email{fake_email_fix_later@fake-email.com}  

\author[orcid=0009-0008-1839-2969]{Fabian Hervas Peters}
\affiliation{Universit\'e Paris Cit\'e, Universit\'e Paris-Saclay, CEA, CNRS, AIM, 91191,
Gif-sur-Yvette, France}
\email{fake_email_fix_later@fake-email.com}  

\author[orcid=0000-0002-9814-3338]{Hendrik Hildebrandt}
\affiliation{Ruhr University Bochum, Faculty of Physics and Astronomy, Astronomical
Institute (AIRUB), German Centre for Cosmological Lensing, 44780 Bochum}
\email{fake_email_fix_later@fake-email.com}  

\author[orcid=0000-0001-9513-7138]{Martin Kilbinger}
\affiliation{Universit\'e Paris Cit\'e, Universit\'e Paris-Saclay, CEA, CNRS, AIM, 91191,
Gif-sur-Yvette, France}
\email{fake_email_fix_later@fake-email.com}  

\author[orcid=0000-0003-4666-6564]{Alan W.~McConnachie}
\affiliation{National Research Council of Canada, Herzberg Astronomy \& Astrophysics
Research Centre, 5071 West Saanich Road, Victoria, BC V9E 2E7, Canada}
\email{fake_email_fix_later@fake-email.com}  

\author[orcid=0000-0002-2637-8728]{Ludovic van Waerbeke}
\affiliation{Department of Physics and Astronomy, University of British Columbia,
Vancouver, BC, V6T 1Z1, Canada}
\email{fake_email_fix_later@fake-email.com}  

\author[orcid=0000-0002-8173-3438]{Anna Wittje}
\affiliation{Ruhr University Bochum, Faculty of Physics and Astronomy, Astronomical
Institute (AIRUB), German Centre for Cosmological Lensing, 44780 Bochum}
\email{fake_email_fix_later@fake-email.com}  


\begin{abstract}
We use galaxy-galaxy lensing to investigate how the dark matter (DM) haloes and stellar
content of galaxies with $0.012 \leq z \leq 0.32$ and $10 \leq
\log_{10}(\Mstar/\mathrm{M}_\odot) \leq 12$ change as a result of the merger process. To
this end, we construct two samples of galaxies obtained from the Ultraviolet Near Infrared
Optical Northern Survey (UNIONS), comprising 1\,623 post-mergers and $\sim$30\,000
non-merging controls, that live in low-density environments to use as our lenses. These
samples are weighted to share the same distributions of stellar mass, redshift, and
geometric mean distance to a galaxy's three nearest neighbours to ensure differences in
the lensing signal are due to the merger process itself. We do not detect a statistically
significant difference in the excess surface density profile of post-mergers and
non-merging controls with current data. Fitting haloes composed of a point-like stellar
mass component and an extended DM structure described by a Navarro-Frenk-White profile to
the lensing measurements yields, for both samples, halo masses of $\mhalo \sim
4\times10^{12}\Msun$ and a moderately negative correlation between \mhalo and
concentration \c. This allows us to rule out, at the 95\% confidence level, merger-induced
starbursts in which more than 60\% of the stellar mass is formed in the burst. The
application of our methods to upcoming surveys that are able to provide samples
$\sim$10$\times$ larger than our current catalogue are expected to detect the weak-lensing
signatures of mergers and further constrain their properties.
\end{abstract}




\section{Introduction}\label{sec:intro}

Amongst the various drivers of galaxy evolution, perhaps the most spectacular is the rapid
transformation of a galaxy's physical and morphological properties due to galaxy mergers
(e.g., via tidal forces, mass growth, gas flows, etc.). On longer timescales, in-situ star
formation (SF) gradually alters galaxies by creating, expelling, and reconstituting new
elements, forming the great diversity of galaxy properties and stellar populations seen
today. These two mechanisms are not independent, however. Galaxy mergers are known to
trigger bursts of SF \citep[e.g.,][]{Sanders1988, Duc1997, Jogee2009, Ellison2013}, but
they can equally quench SF on short timescales \citep[e.g.,][]{Ellison2022, Davies2022,
Ellison2024, Kado-Fong2024}. The interplay of these two processes complicates our
understanding of the long-term effects that mergers may have on SF in galaxies.

While star formation rates (SFRs) in mergers are widely studied
\citep[e.g.,][]{Robaina2009, Ellison2013, Knapen2015, Moreno2019, Pearson2019b, Shah2022},
the total stellar mass produced as a result of merging is harder to probe since it
requires integrating SFRs over many dynamical times, and is thus less well-understood.
Recent results from \citet{Reeves2024} and \citet{Ferreira2024_oct}, each using different
techniques, both show that the SF triggered by the merger process is responsible for
10--20\% of the post-merger galaxy's stellar mass. This, in turn, changes the
stellar-to-halo-mass ratio (SHMR) of the post-merger. Thus, measuring a galaxy's SHMR is
another way to understand how many stars could have formed in a merger-induced starburst.
To demonstrate this, we consider the simple case of a merger between two galaxies sharing
equal halo (\mhalo) and stellar (\Mstar) masses, forming a new system with $\Mstar \approx
10^{11} \Msun$. Using the fits to the SHMR as a function of \Mstar in \cite{Hudson2015},
we see that if the merger was at $z=0.15$ and was ``dry'' (i.e., no induced SF), it would
have a SHMR $\approx\,$25\% higher than a non-merging galaxy of the same stellar mass. If,
on the other hand, the merger triggers a $\sim$20\% starburst (i.e., the fraction of stars
formed due to merging is 20\% of the post-merger's final stellar mass), the SHMR
immediately after the burst would be $\approx\,$60\% higher than a galaxy of the same
\Mstar.

In addition to changing stellar mass and SHMR, galaxy interactions are also expected to
change the mass density profile of merging systems and their products \citep{Wang2020},
which we can measure using the weak gravitational lensing signal of background galaxies
around the positions of the mergers---that is, using galaxy-galaxy lensing. The mass
profile of mergers can, of course, be studied using simulations
\citep[e.g.,][]{Springel2005, Hopkins2006}, but simulating the baryonic content of
galaxies is very computationally challenging while modelling a galaxy's dark matter is
much simpler. Conversely, it is impossible to observe the dark matter (DM) directly,
whereas the baryonic matter is observable via electromagnetic radiation. Using
galaxy-galaxy lensing, we can probe the DM content of observationally-identified mergers
and thus act as a bridge between theory and observation.

Until recently, however, a sufficiently large catalogue of merger products was simply
unavailable for the purposes of a weak lensing analysis. While pre-mergers (i.e., pairs
that will soon merge or systems that have started merging but have not fully coalesced)
are relatively easy to find based on just position and redshift \citep[see,
e.g.,][]{Ellison2008, Domingue2009, Xu2012, Gonzalez2023}, identification of post-mergers
(i.e., systems that have fully coalesced after a merger) is much harder
\citep{Wilkinson2024} and manual inspection cannot keep up with the meteoric growth in
astronomy data sizes. Fortunately, machine learning provides a promising avenue for
automated merger identification using neural networks (NNs).

Historically, convolutional neural networks (CNNs) have been the NN of choice for
identifying galaxy mergers due to their exceptional performance in image processing and
analysis \citep[e.g.,][]{HuertasCompany2015, Pearson2019, Pearson2019b, Bickley2021,
Rose2024}. Recently, \citet{Ferreira2024} combined vision transformers (ViTs) with CNNs in
a two-stage hierarchical merger identification framework, achieving post-merger purities
$\gtrsim\,$95\% even at $\sim$1\Gyr after the merger event.

Motivated by the success of NNs in identifying mergers and the release of new, large
catalogues of candidate mergers suitable for a weak lensing analysis, here we investigate
how galaxy mergers change the DM structure and stellar content of galaxies using
galaxy-galaxy lensing. Previously, the only instance that we know of that attempted to use
galaxy-galaxy lensing to study galaxy mergers (albeit in the context of mergers triggering
black hole activity, for which they found a null result) is from \citet{Harvey2015}, who
looked at 29 post-starburst galaxies containing quasars whose SF histories indicated
recent merger activity. In this work, we increase the sample size of post-mergers by a
factor of $\approx\,$56 and we do not rely on the post-starburst signature to identify
merger products, since post-starburst mergers comprise a minority of post-merger systems
(\citeauthor{Ellison2022} \citeyear{Ellison2022} find $\lesssim\,$20\% of post-mergers are
also post-starbursts), meaning the properties of post-starburst galaxies may not be
reflective of post-mergers as a whole.

Throughout this paper, we assume a flat $\Lambda$CDM cosmology with
$H_0=70\km\s^{-1}\Mpc^{-1}$, $\Omega_{\mathrm{m},0}=0.3$, $\Omega_{\Lambda,0}=0.7$,
$\Omega_{\mathrm{b},0}=0.049$, $\sigma_8=0.81$, and $n_\mathrm{s}=0.95$. We outline the
data used for our project in Section \ref{sec:data}, followed by our methods in Section
\ref{sec:methods}. We then present and discuss the results of our analysis in Section
\ref{sec:results}, and conclude in Section \ref{sec:conclusions}.

\section{Data}\label{sec:data}

The data for our analysis consist of a lens catalogue selected from the
\citet{Ferreira2024} dataset comprising galaxies at various stages in the merger sequence,
and a background (source) galaxy shape catalogue. Both of these are derived from the
Ultraviolet Near Infrared Optical Northern Survey \citep[UNIONS;][]{Gwyn2025} $r$-band
images. UNIONS is a collaboration of wide-field imaging surveys of the northern
hemisphere, consisting of the Canada-France Imaging Survey (CFIS) conducted at the 3.6\m
Canada-France-Hawaii Telescope (CFHT) on Maunakea, members of the Pan-STARRS (Panoramic
Survey Telescope and Rapid Response System) team, and the Wide Imaging with Subaru Hyper
Suprime-Cam of the Euclid Sky (WISHES) team. CFHT/CFIS is obtaining deep imaging in the
\textit{u}- and \textit{r}-bands, Pan-STARRS is obtaining deep \textit{i}- and
moderate-deep \textit{z}-band imaging, and Subaru is obtaining deep imaging in the
\textit{z}-band through WISHES and \textit{g}-band through the Waterloo-Hawaii Institute
for Astronomy \textit{g}-band Survey (WHIGS). In the following, we describe our two
datasets in more detail.

\begin{figure*}
  \centering
  \begin{subfigure}[t]{0.33\textwidth}
    \vspace*{0pt}
    \centering
    \includegraphics[width=\textwidth]{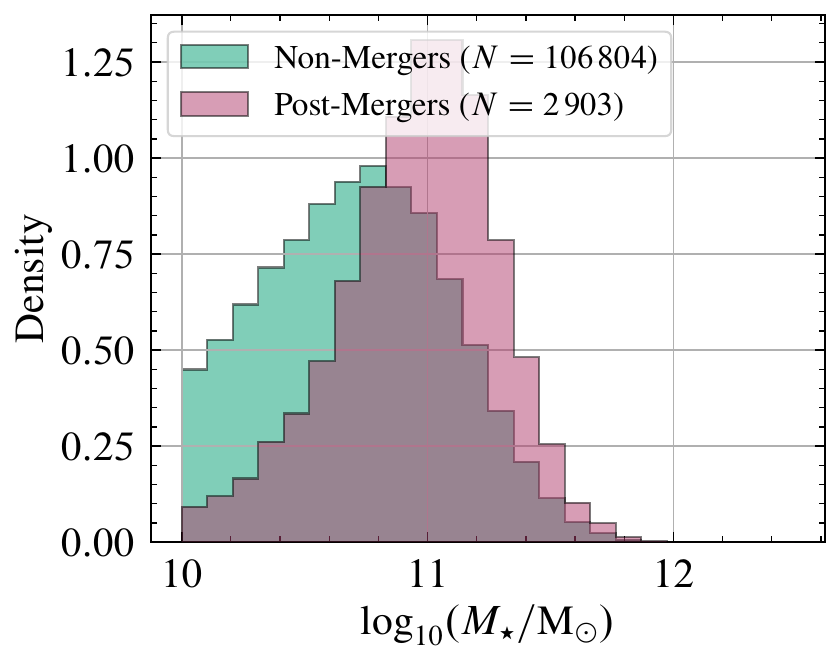}
    \caption{}
    \label{subfig:Mstar_distr_unmatched}
  \end{subfigure}
  \hfill
  \begin{subfigure}[t]{0.33\textwidth}
    \vspace*{0pt}
    \centering
    \includegraphics[width=\textwidth]{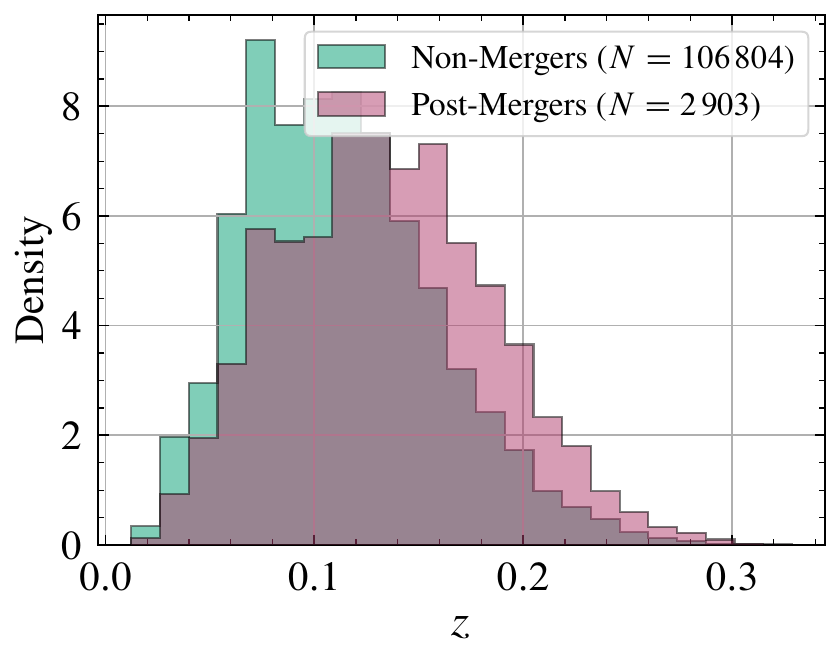}
    \caption{}
    \label{subfig:z_distr_unmatched}
  \end{subfigure}
  \hfill
  \begin{subfigure}[t]{0.33\textwidth}
    \vspace*{0pt}
    \centering
    \includegraphics[width=\textwidth]{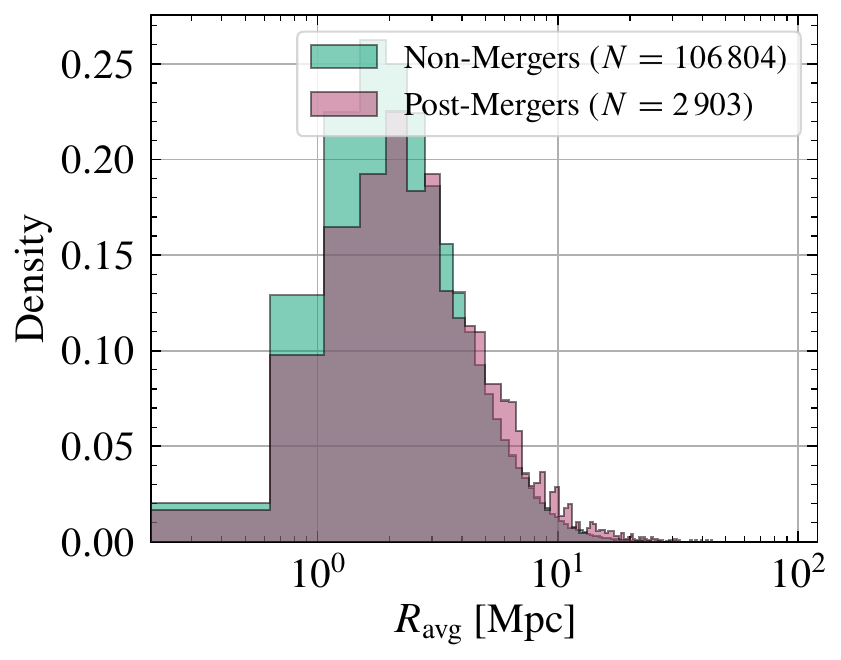}
    \caption{}
    \label{subfig:Ravg_distr_unmatched}
  \end{subfigure}
  \caption{Raw distributions of various galaxy properties in our lens catalogue ($\Mstar
  \geq 10^{10}\Msun$), before matching sample properties.
  \hyperref[subfig:Mstar_distr_unmatched]{(a)} The raw stellar mass distribution.
  \hyperref[subfig:z_distr_unmatched]{(b)} The raw redshift distribution.
  \hyperref[subfig:Ravg_distr_unmatched]{(c)} The raw distribution of the geometric mean
  distance to each galaxy's three nearest neighbours.}
  \label{fig:all_distr_unmatched}
\end{figure*}

\subsection{Lens Galaxies}\label{sec:lenses}

To study the DM haloes of galaxy mergers, we use the catalogue from \citet{Ferreira2024}
that includes 235\,354 galaxies at different stages in the merger
sequence:~pre-mergers/pairs, post-mergers, and non-mergers. The objects in this catalogue
are composed of galaxies with spectroscopic data in the Sloan Digital Sky Survey (SDSS)
Data Release 7 (DR7) that are additionally contained in the UNIONS Data Release 5 (DR5)
survey area. We do not use the pre-mergers/pairs in our subsequent analysis because we do
not expect them to be well-described by a single DM halo, and is thus beyond the scope of
this paper.

Each galaxy in our catalogue has an SDSS DR7 object ID, (RA, Dec) on-sky coordinate, the
vote count from each of the two \textsc{Mummi} NN stages (described below), and a ``merger
class'' designation indicating whether the galaxy is a non-merger, part of a pre-merger
pair, or is a post-merger. We additionally use the object IDs to match galaxies to SDSS
and MPA-JHU (Max Planck Institute for Astrophysics and Johns Hopkins University)
catalogues to obtain, for each of the 184\,485 galaxies in common across all three
datasets, a spectroscopic redshift \z from SDSS and a stellar mass estimate \Mstar from
MPA-JHU\footnote{\url{https://wwwmpa.mpa-garching.mpg.de/SDSS/DR7/Data/stellarmass.html}}.
The stellar masses are derived from SDSS \textit{ugriz} photometry using the template
models of \citet{Bruzual2003} assuming a \citet{Chabrier2003} initial mass
function\footnote{Note that these \Mstar estimates are similar, but not identical, to
those of \citet{Kauffmann2003}, as stated in the MPA webpage linked in the previous
footnote. For a more detailed comparison between the MPA-JHU stellar masses and those in
\citet{Kauffmann2003}, see
\url{https://wwwmpa.mpa-garching.mpg.de/SDSS/DR7/mass_comp.html}.}.

Merger identification and merger stage classification are based on the values given by
\textsc{Mummi} \citep{Ferreira2024}, a novel NN ensemble composed of convolutional and
vision transformer units. We summarize the NN training, architecture, and results below.
The \textsc{Mummi} training and validation data come from mock images of IllustrisTNG100-1
galaxies \citep{Nelson2019} spanning $0.01 \leq z \leq 0.3$ and $10^{10}\Msun \leq \Mstar
\leq 10^{11}\Msun$ to generally replicate the morphologies seen in UNIONS. To generate
these data, 17\,894 pre-mergers and 21\,485 post-mergers are selected from TNG100-1; the
latter are galaxies whose progenitors have stellar mass ratios $\mu\geq0.1$ and have
merged in the last 1.7\Gyr, with no companions within 50\kpc. Each merger is matched in
redshift, stellar mass, and gas fraction to a non-merging control from TNG100-1. A
non-merger is a galaxy that has no merger event within 1.7\Gyr, both in the past and in
the future (relative to its observed redshift). Non-mergers that could potentially merge
in the future after the simulation ends have a minimum separation of 50\kpc so that they
are unlikely to merge within the next Gyr \citep{Patton2024}. Then, for each galaxy, 80
mock UNIONS realizations are produced using \textsc{RealSimCFIS}, a pipeline based on
\textsc{RealSim} \citep{Bottrell2019, Bottrell2024}.

The 6\,300\,640 mock images are classified by \textsc{Mummi} in a two-step process:~first
determining if a galaxy is a merger or non-merger, then separating the mergers into
pre-mergers and post-mergers. Step 1 (merger identification) is accomplished with an
ensemble of 20 NNs (10 pairs of CNNs and SwinTransformers), with a positive merger
identification only if all 20 models agree. Step 2 (merger classification) uses a single
CNN-SwinTransformer pair to determine the probability that the galaxy is a post-merger,
with a ``merger class'' label of post-merger if the probability is greater than 50\%. This
hierarchical approach leads to samples that are above 95\% purity for post-mergers that
are observed within 1.25\Gyr of the merger event.

The trained \textsc{Mummi} ensemble is applied to UNIONS/CFIS \textit{r}-band cutouts with
a 12 Petrosian radius field of view, giving us 3\,096 galaxies unanimously deemed to be
post-mergers in the UNIONS/SDSS DR7/MPA-JHU overlap. These post-mergers are generally
isolated and feature disturbed morphologies such as tidal tails and warped spiral arms
\citep[for more details, see][]{Ferreira2024}. The requirement of unanimity in
\textsc{Mummi} step 1 maximizes the purity of our post-merger sample. Likewise, to
increase purity in the non-merging sample, a non-merging candidate may have no more than
three votes classifying it as a merger in \textsc{Mummi} step 1; if we simply classified
non-mergers as those that do not have unanimous agreement in the first step, there will be
many false positives such as a merging system that is correctly identified by (only) 19 of
the 20 models. We note that, even with this scheme, it is possible that there is a small
number of false positives among the 132\,790 non-mergers, but the contaminating fraction
should be small on account of the rarity of mergers in the Universe \citep[see,
e.g.,][]{Casteels2014, Ferreira2024} and these missed mergers should be dominated by post-
rather than pre-mergers. This is because pre-mergers, in contrast to post-mergers, are
actually fairly easy to identify (see also the discussion in Section \ref{sec:intro}).
Thus, there is little chance that a pre-merger only attains 3 votes or fewer in the first
step of \textsc{Mummi}, meaning any false positives in the non-merging sample are likely
to be post-mergers. We note here that, as discussed in \citet{Ferreira2024}, the
\textsc{Mummi} post-merger galaxies tend toward higher redshifts and stellar masses than
the overall sample of galaxies in SDSS/UNIONS (see Figure \ref{fig:all_distr_unmatched}).
In this work, we do not make any attempt to correct for this selection effect.

Lastly, we impose a minimum stellar mass of $10^{10}\Msun$ in our samples to maximize the
confidence in our galaxy classifications, as \textsc{Mummi} is designed for galaxies above
this stellar mass threshold\footnote{Note, however, that \textsc{Mummi} classifications
should generalize to stellar masses outside of its training range \citep[section
4.4]{Ferreira2024}. As we discuss in the text, our results are robust to this \Mstar
cut.}. In contrast, we do not need to apply additional redshift cuts since our samples
(Figure \ref{fig:all_distr_unmatched}) naturally span a redshift range of $0.012 \leq z
\leq 0.33$, which ensures our galaxies and any close neighbours fit within the 12
Petrosian radius field of view used by \textsc{Mummi} and minimizes the chance that an
unresolved point-like source, such as a quasar, is accidentally included in our sample
\citep{Ferreira2024}. The addition of a stellar mass cut on our samples does not change
any of the results presented in this paper; for reference, our \Mstar cut reduces our
final post-merger and non-merging control sample sizes (see Section \ref{sec:methods}) by
4\% and 6\%, respectively.

We show in Figures \ref{subfig:Mstar_distr_unmatched} and \ref{subfig:z_distr_unmatched}
the distributions of stellar mass and redshift in our samples. In order to determine the
environment of the lenses, we also convert the on-sky coordinates of each galaxy into a
three-dimensional position using its spectroscopic redshift and the cosmology given in
Section \ref{sec:intro}, assuming the galaxies follow the Hubble flow without any peculiar
velocities. We then calculate the distance to each galaxy's three nearest
neighbours\footnote{Nearest in three-dimensional space, not nearest in projected 2D on-sky
separation.} and compute the geometric mean of these three separations, which we call
\ravg; we use these values in Section \ref{sec:env_control}. Figure
\ref{subfig:Ravg_distr_unmatched} shows the resulting \ravg distribution.

\subsection{Background Source Galaxies}\label{sec:sources}

The shapes of the background source galaxies come from the UNIONS ShapePipe v1.3
catalogue, a collection of $\sim$84 million galaxies with shape measurements derived from
CFIS \textit{r}-band data. These sources are stacked in annuli around each of our lenses
to perform the tangential shear measurement described in Section \ref{sec:measurements}.
In the following, we provide a brief overview of the shape measurement pipeline
\citep{Guinot2022, Farrens2022} including v1.3 improvements \citep{Li2024}.

Objects, including both point-like and extended sources, are initially extracted from
stacked \textit{r}-band CFIS images using \textsc{SourceExtractor} \citep{Bertin1996} with
a 1.5$\sigma$ detection threshold. The minimum area to detect an object in v1.3 has been
changed from 10 to 3 pixels to reduce galaxy selection bias in the resulting shear
estimates. Next, to differentiate between point sources and galaxies, each object is
compared to a ``spread model'' \citep{Desai2012, Mohr2012} consisting of a point source
profile and an extended profile. An object is considered a galaxy if its spread model is
greater than zero while satisfying a number of other cuts based on the model uncertainty,
object magnitude, size of the extended source relative to the size of the point spread
function (PSF), and the relative error in the flux. In v1.3, estimates of the PSF are
obtained using multi-CCD (MCCD) modelling \citep{Liaudat2021} instead of \textsc{PSFEx}
\citep{Bertin2011}. The MCCD approach gives a simultaneous non-parametric estimate of the
PSF across all CCDs to better capture PSF variations over the whole field of view.
Moreover, the size of the galaxy ($T_\text{gal}$) relative to the PSF ($T_\text{PSF}$) has
an additional constraint in v1.3 to reject diffuse, low signal-to-noise artifacts in the
image:~$T_\text{gal}/T_\text{PSF}<3$. Lastly, galaxy shapes are measured using
\textsc{ngmix}\footnote{\url{https://github.com/esheldon/ngmix}} and calibrated using the
metacalibration method \citep{Huff2017, Zuntz2018}.

\begin{figure}
  \centering
  \includegraphics[width=\columnwidth]{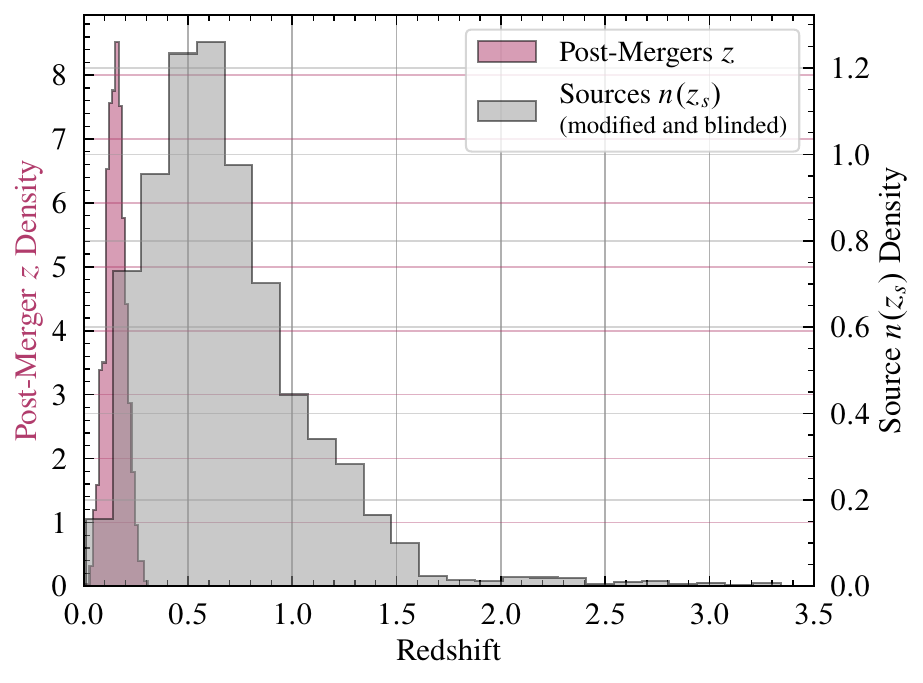}
  \caption{The effective redshift distribution of the sources, $n(z_s)$, compared to the
  post-merger spectroscopic redshift distribution from Figure
  \ref{subfig:z_distr_matched}. For cosmology calculations, the $n(z_s)$ redshifts are
  still blinded; the source redshift distribution in this plot is close, but not
  identical, to the source redshift catalogue used in this paper.}
  \label{fig:nz}
\end{figure}

Notably, there are no spectroscopic redshifts for these sources at present, but the
effective redshift distribution $n(z_s)$ for the photometric sample is estimated from
colour space based on the method from \citet{Wright2020}. The process applied to UNIONS is
described in \citet[appendix A]{Li2024} and we summarize it here. The sources are first
cross-matched to the CFHTLenS \citep[Canada-France-Hawaii Telescope Lensing
Survey,][]{Heymans2012} W3 field and assigned \textit{ugriz} magnitudes
\citep{Hildebrandt2012}. These magnitudes and colours are input to a self-organizing map
\citep[SOM,][]{Kohonen1982} to reduce their multi-dimensional space to a 2D plane. A
sample of galaxies with the same photometry and spectra from external surveys are also
input to the SOM as a calibration dataset, and their redshift distribution is weighted
such that it matches that of the total source sample, based on the lensing weights and the
number density per SOM region. The resulting weighted distribution is then used as an
estimate of the photometric sample's $n(z_s)$, as shown in Figure \ref{fig:nz}. This
$n(z_s)$ distribution is used to calculate the effective inverse critical surface density
$\langle\Sigma^{-1}_\text{crit}\rangle$, described in Section \ref{sec:measurements}.

As a final note, the ShapePipe v1.3 catalogue does not have an estimate for the scalar
bias needed to compute a shear bias correction to our lensing measurements in Section
\ref{sec:measurements}. However, since the bias affects lensing amplitudes by just a few
percent, it is accounted for in our uncertainties, which are all substantially larger (see
Section \ref{sec:results}).

\section{Methods}\label{sec:methods}

We now describe the procedure for our weak lensing analysis. We prepare our lenses (i.e.,
the post-mergers as well as the non-merging galaxies from which we will select our
controls) for lensing measurements in Sections \ref{sec:env_control} and
\ref{sec:weighting} by ensuring their redshift and stellar mass distributions match, as
well as by controlling our sample so that the galaxies all live in similar, low-density
environments. We always match our distributions to the post-merger sample, since that is
our smallest dataset and we do not want to further reduce its effective size. These
preprocessing steps are important because they ensure that we are measuring differences in
the lensing signal caused by the merger process, as opposed to differences caused by
sampling separate galaxy populations with distinct, merger-unrelated properties or
environments.

In addition to matching the lens samples' \Mstar, \z, and \ravg distributions, we also
construct a set of random lenses in Section \ref{sec:randoms} that we use for various
lensing correction factors. Section \ref{sec:measurements}, in turn, describes the
computational method by which we perform the lensing measurements. Finally, Section
\ref{sec:fitting} details the procedure for fitting DM haloes to the lensing signal.

\subsection{Environment Control}\label{sec:env_control}

\begin{figure*}
  \centering
  \includegraphics[width=\textwidth]{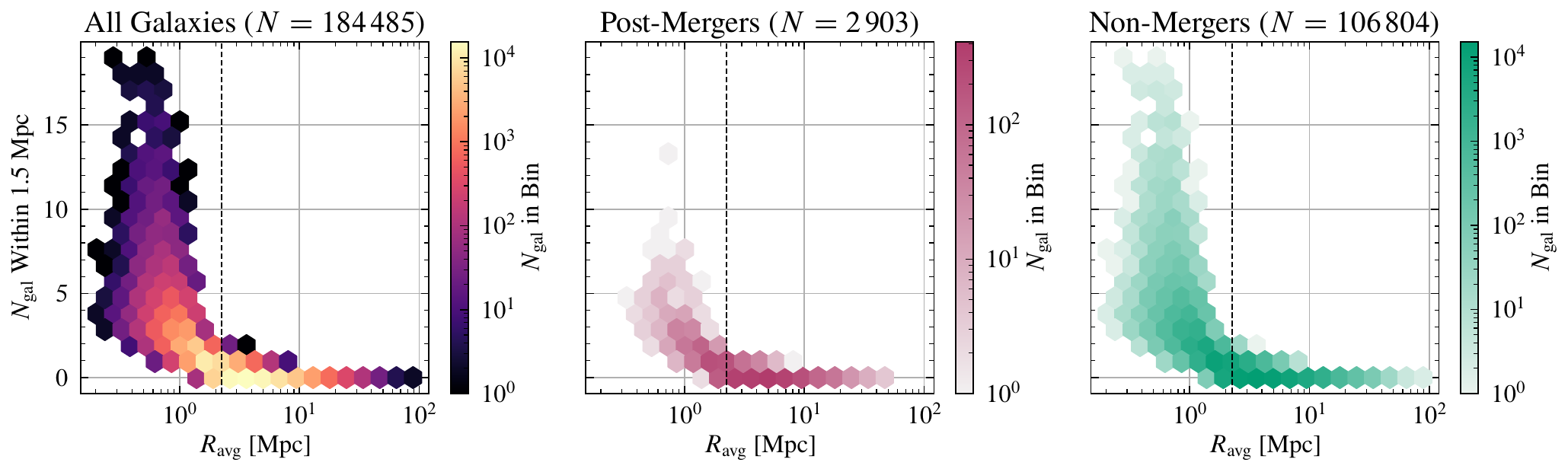}
  \caption{2D hexbin plots showing our metrics for environment density:~the number of
  neighbouring galaxies within 1.5\Mpc and the geometric mean distance to the three
  nearest neighbours of each galaxy, \ravg. The colouring represents the number of
  galaxies, $N_\mathrm{gal}$, in a hexbin. The vertical dashed line indicates $\ravg =
  2.25\Mpc$. We want to select galaxies that live in low-density environments (i.e., the
  bottom right region of these plots).}
  \label{fig:envNumber_2d_hexbin}
\end{figure*}

We begin by preferentially choosing lenses that live in low-density environments to
mitigate the need to model the correlation between our lens galaxies and neighbouring
haloes \citep[the so-called 2-halo term, see e.g.][]{Cooray2002}. This also ensures that
the post- and non-merger lens samples contain galaxies that all live in similar
environments.

We determine the extent to which a galaxy is isolated using two metrics:~the number of
other galaxies within a distance of 1.5\Mpc (i.e., the upper limit of our radial bins in
our lensing measurements, see Section \ref{sec:measurements}), and the geometric mean
distance to a galaxy's three nearest neighbours. Distances are calculated from a galaxy's
spectroscopic redshift converted into Cartesian coordinates using its on-sky location and
angular diameter distance, assuming the cosmology from Section \ref{sec:intro} and
neglecting possible peculiar velocities on top of the Hubble flow. To compute these
metrics, we simply search within the catalogue of 184\,485 galaxies described in Section
\ref{sec:lenses} as these data come from a complete sample of spectroscopic redshifts and
alternative samples would not have full overlap with our footprint. Note that these
searches consider \emph{all} galaxies in the catalogue including pre-mergers and galaxies
with $\Mstar < 10^{10}\Msun$, not just the post-mergers and non-mergers that we use.

Since we are using spectroscopic redshifts from SDSS DR7, there may be neighbouring
galaxies that are missed in these searches due to fibre collisions. These fibre collisions
are an instrument limitation where the fibres on the spectrographic plates used by SDSS
cannot be placed closer than 55\arcsec \citep{Strauss2002}; galaxies closer than 55\arcsec
in separation are selected at random\footnote{Since all galaxies, excluding those hosting
candidate quasars, have equal priority in the SDSS spectroscopic target selection
\citep[see][section 4.5]{Strauss2002}.} to be observed spectroscopically
\citep{Strauss2002}. Although only $\sim$6\% of galaxies in SDSS lack spectra due to the
physical limitations of fibre placement \citep{Strauss2002, Tago2010}, \citet{Patton2008}
find that $\lesssim\,$30\% of pairs with projected angular separation $<\,$55\arcsec have
spectroscopic redshifts for both galaxies\footnote{This 30\% completeness number also
includes false pairs that are only close in projection. The completeness fraction for true
pairs (i.e., those close in 3D space) with spec-$z$ relative to all true pairs should be
higher.}. However, \textsc{Mummi} results show that the unanimous post-mergers in
UNIONS/CFIS are generally isolated galaxies \citep{Ferreira2024}, so our samples (after
matching the non-mergers to the post-mergers in Section \ref{sec:weighting}) should indeed
live in relatively low-density environments. Furthermore, while having a low-density
environment makes modelling DM haloes simpler, it is not necessary for our lensing
measurements, which only require that the galaxy samples live in the \emph{same}
environment---this we achieve with our strategy presented below.

Two-dimensional histograms of our two environmental density metrics are shown in Figure
\ref{fig:envNumber_2d_hexbin}. To select lenses that live in low-density environments, we
choose galaxies with no companions within 1.5\Mpc. Additionally, we also require that the
average distance to a galaxy's three nearest neighbours, $\ravg$, be greater than
1.5$\times$ the 1.5\Mpc value. This is because the centres of some neighbouring galaxies
may be just outside the 1.5\Mpc search radius, rendering them invisible to our first
metric, but the DM haloes of these neighbouring galaxies will still influence lensing
amplitudes in the upper radial bin ending at 1.5\Mpc. Hence, we conservatively require the
minimum \ravg to be greater than 2.25\Mpc.

\subsection{Weighting of Lenses}\label{sec:weighting}

\begin{figure*}
  \centering
  \begin{subfigure}[t]{0.33\textwidth}
    \vspace*{0pt}
    \centering
    \includegraphics[width=\textwidth]{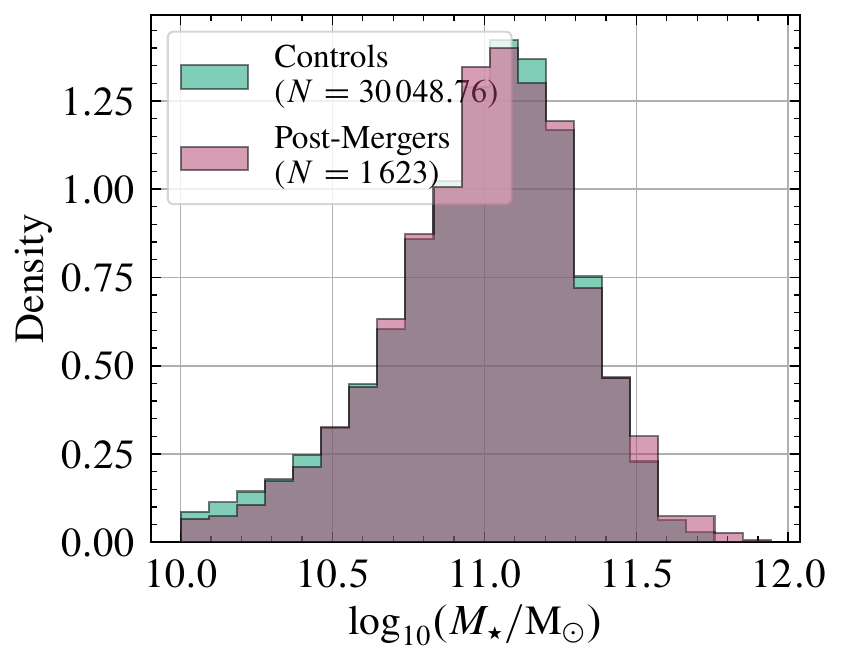}
    \caption{}
    \label{subfig:Mstar_distr_matched}
  \end{subfigure}
  \hfill
  \begin{subfigure}[t]{0.33\textwidth}
    \vspace*{0pt}
    \centering
    \includegraphics[width=\textwidth]{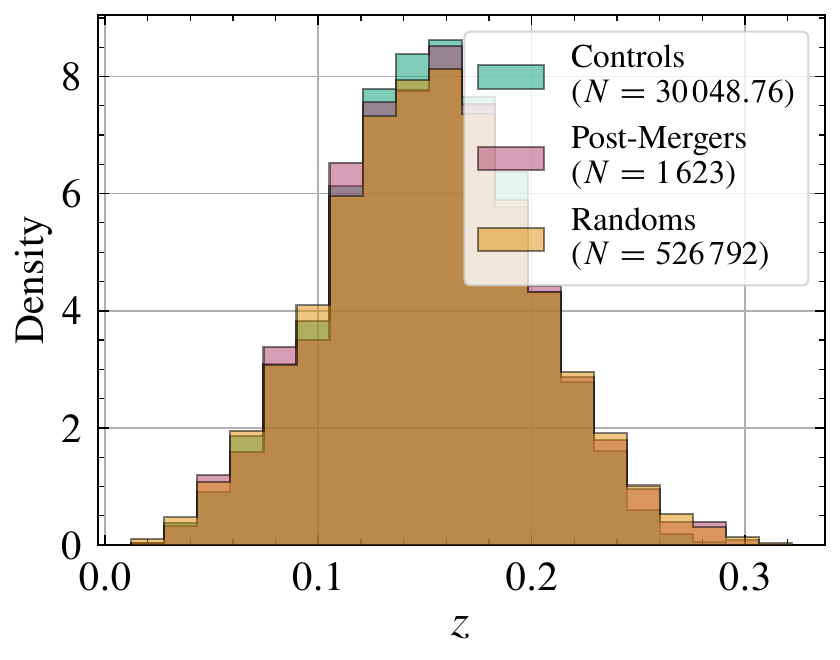}
    \caption{}
    \label{subfig:z_distr_matched}
  \end{subfigure}
  \hfill
  \begin{subfigure}[t]{0.33\textwidth}
    \vspace*{0pt}
    \centering
  \includegraphics[width=\textwidth]{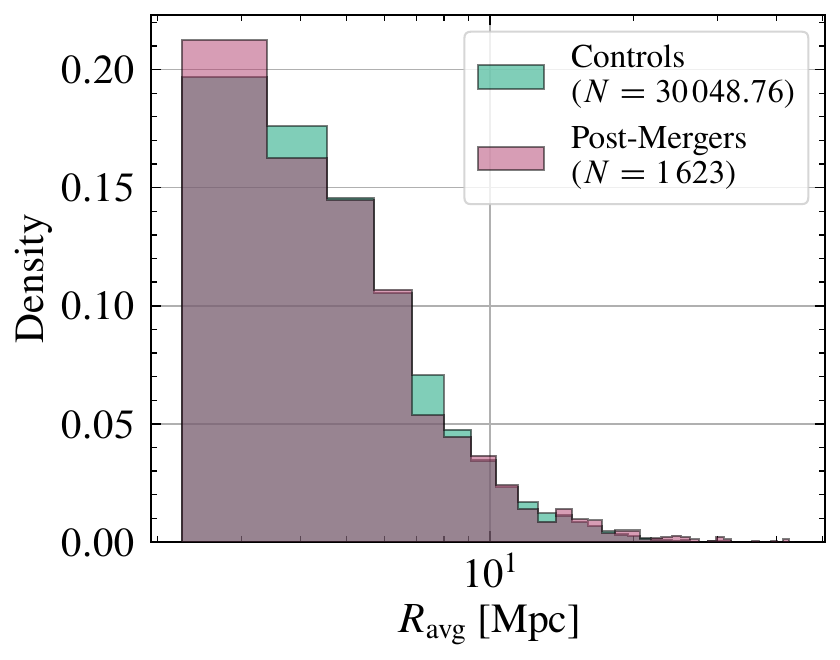}
    \caption{}
    \label{subfig:Ravg_distr_matched}
  \end{subfigure}
  \caption{The weighted distributions of our final lens samples' galaxy properties. These
  galaxies are selected to live in low-density environments, as discussed in the text, and
  matched in \hyperref[subfig:Mstar_distr_matched]{(a)} stellar mass,
  \hyperref[subfig:z_distr_matched]{(b)} redshift, and
  \hyperref[subfig:Ravg_distr_matched]{(c)} the geometric mean distance to their three
  nearest neighbours. Two-sample KS test $p$-values for the controls and post-mergers
  are:~\hyperref[subfig:Mstar_distr_matched]{(a)} 0.38,
  \hyperref[subfig:z_distr_matched]{(b)} 0.35, and
  \hyperref[subfig:Ravg_distr_matched]{(c)} 0.34, indicating that differences in these
  weighted samples are not statistically significant. The redshifts of the randoms come
  from directly sampling the post-merger's redshift KDE; a KS test between these two
  distributions yields $p=0.80$.}
  \label{fig:all_distr_matched}
\end{figure*}

After rejecting galaxies that do not meet the two conditions in Section
\ref{sec:env_control}, we fit three-dimensional kernel density estimates (KDEs) to the
resulting samples' joint \Mstar-\z-\ravg distributions (i.e., one joint distribution for
post-mergers and one joint distribution for non-mergers). The KDEs are used to assign
weights to the non-merger sample according to the formula:
\begin{equation}\label{eq:weights}
  w = \frac{f_\mathrm{post}(\Mstar, \z, \ravg)}{f_\mathrm{non}(\Mstar, \z, \ravg)},
\end{equation}
where $w$ is the weight given to a non-merging galaxy, $f_\mathrm{post}$ and
$f_\mathrm{non}$ are the post- and non-merger KDEs, and the \Mstar, \z, \ravg values refer
to the sample being weighted (i.e., the non-mergers). These weights ensure the
distributions between our two samples are the same, as seen in Figure
\ref{fig:all_distr_matched}. In this way, we have constructed a non-merging control sample
that is matched to the post-mergers. Two-sample Kolmogorov-Smirnov (KS) tests performed on
the distributions verify that the post-mergers and controls have similar properties
($p$-values of 0.38, 0.35, 0.34 for the stellar mass, redshift, and \ravg distributions,
respectively).

Note that we choose to fit KDEs to the data instead of directly assigning weights from
histogram bins because histograms can be very jagged, but we expect the underlying
distributions of \Mstar, \z, and \ravg to be smooth and continuous\footnote{Depending on
the histogram binning, galaxies on the edge of two bins could drastically change their
weight just by falling into a different bin. This will not happen with KDEs since they are
continuous distributions where small changes in inputs lead to small changes in outputs.}.
Thus, fitting KDEs to the data allows us to obtain smoothly varying weights over
continuous intervals.

The final effective sample sizes are:~1\,623 post-mergers and about 30\,000 non-merging
controls. An effective sample size (ESS) accounts for possibly unequal weighting of data
and is calculated as \citep[equation (4.3)]{Kish1992}:
\begin{equation}\label{eq:kish}
  \text{ESS} = \frac{\left(\sum_i w_i\right)^2}{\sum_i w_i^2},
\end{equation}
where $w_i$ are the weights of the galaxies in a given sample ($w_i=1$ for all
post-mergers) and the sums run over the entire sample. When all weights are equal, the ESS
is equal to the actual sample size, but the ESS decreases as weights become more
concentrated in fewer data points.

\subsection{Construction of Random Lenses}\label{sec:randoms}

\begin{figure*}
  \centering
  \includegraphics[width=0.49\textwidth]{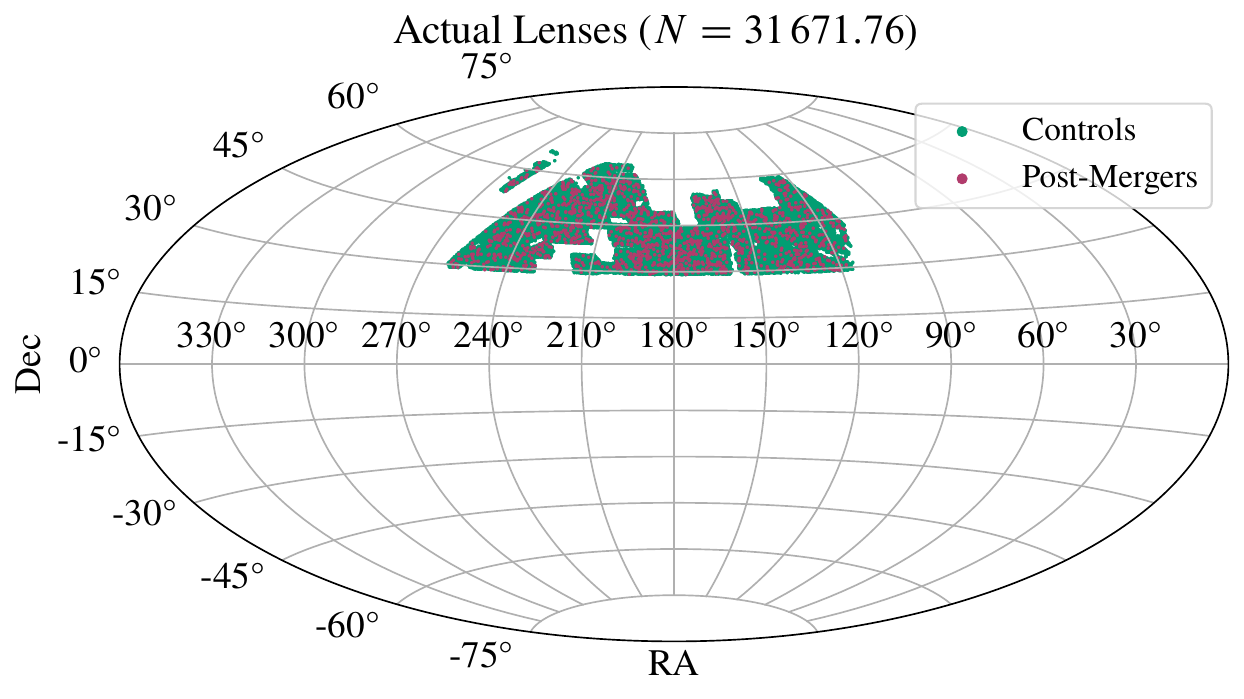}
  \includegraphics[width=0.49\textwidth]{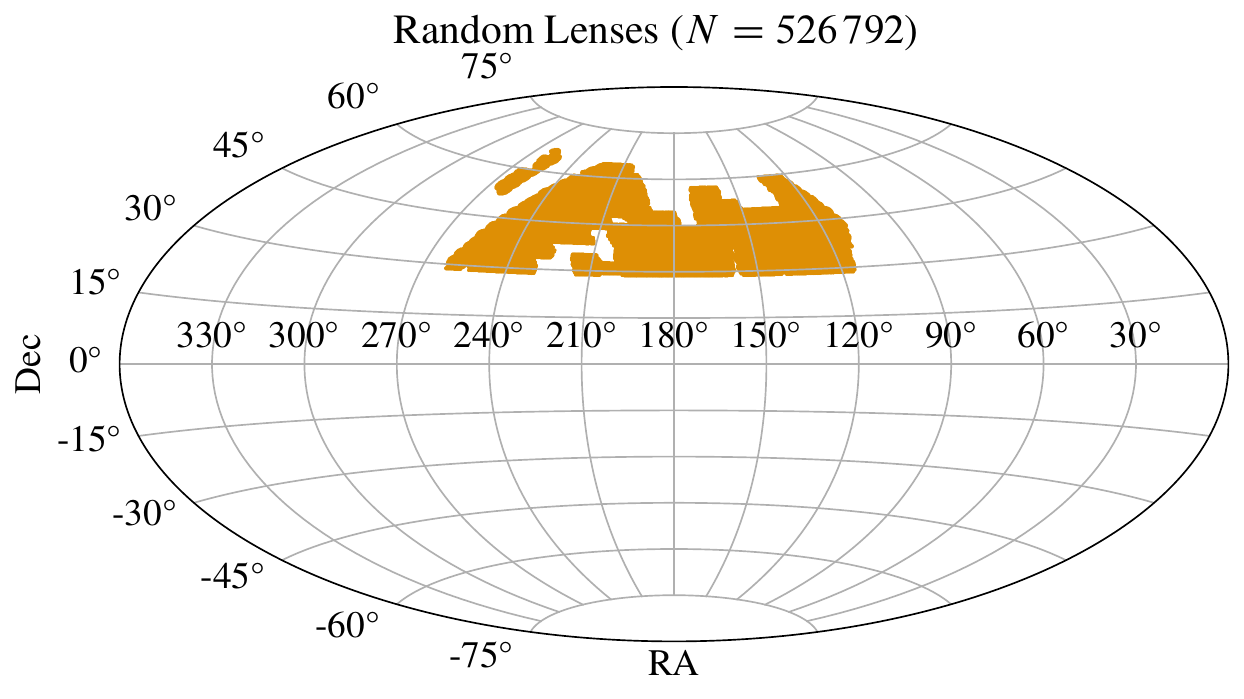}
  \caption{Sky coverage of the true lenses (\textit{left}) and random lenses
  (\textit{right}).}
  \label{fig:sky_coverage}
\end{figure*}

In addition to the photometric redshift correction $n(z_s)$ mentioned in Section
\ref{sec:sources}, we also construct a random catalogue of lenses to correct for two other
effects:~lensing systematics (corrected via random subtraction), and an overabundance of
unlensed sources that are physically near lenses, causing the lensing amplitude to be
underestimated (corrected via the boost factor).

To create our randoms catalogue, we generate $\sim$10$\times$ more random lenses than true
lenses\footnote{The ESS of the random lenses, which are equally weighted, happens to be
$\sim$17$\times$ that of the true lenses.} uniformly within the true lenses'
footprint---namely, the overlapping region between the SDSS DR7 footprint and the
currently-surveyed area by UNIONS. We determine the intersection of these two surveys
using the SDSS DR7 spectroscopic coverage
table\footnote{\url{https://classic.sdss.org/dr7/coverage/maindr72spectro.par}}, which
contains the pointings for the 3\degree-diameter plate centres, and a list of UNIONS
footprint tiles each covering an area of $0\fdg5\times0\fdg5$. We do not implement
additional masking, such as for bad pixels, in our random catalogue. The spatial
distributions of our true lenses and random lenses are shown in Figure
\ref{fig:sky_coverage}, covering an area of 1\,995 square degrees.

We match the redshift distribution of our randoms to the post-mergers by randomly
assigning redshifts sampled from a KDE fitted to the post-mergers' \z-distribution. Figure
\ref{subfig:z_distr_matched} shows the redshift distributions of our various lens samples,
including the randoms catalogue.

\subsection{%
  \texorpdfstring{$\boldsymbol{\dsigma}$}{ΔΣ} Measurements%
}\label{sec:measurements}

After pre-processing the data as described above, we measure their lensing amplitudes
using the \textsc{dsigma} Python package \citep{Lange2022}. Briefly, \textsc{dsigma}
measures the excess surface density \dsigma around our lenses, defined as the mean surface
density within a circle of projected radius $R$ minus the mean surface density at the edge
of a circle of projected radius $R$:
\begin{equation}
  \dsigma(R) = \overbar{\Sigma(<\!R)} - \langle\Sigma(R)\rangle.
\end{equation}
It accomplishes this by binning and stacking background source galaxies in annuli around
each of our (possibly weighted) lenses and averaging the tangential shears,
$\gamma_t$, of the sources. Given a cosmology, this yields an estimate of \dsigma:
\begin{equation}\label{eq:shear}
  \dsigma(R) = \frac{\sum_{l\text{-}s} w_{\text{sys},l} w_{l\text{-}s}
  \langle\Sigma^{-1}_\text{crit}\rangle^{-1} \gamma_t}{
  \sum_{l\text{-}s} w_{\text{sys},l} w_{l\text{-}s}
  },
\end{equation}
where $\sum_{l\text{-}s}$ denotes that the sum goes over each lens-source pair,
$w_{\text{sys},l}$ are systematic weights for the lens samples (equal to the weights in
equation \eqref{eq:weights} for controls and unity for the post-mergers), and
$w_{l\text{-}s}$ are weights to minimize shape noise in our stacked signal:
\begin{equation}\label{eq:wls}
  w_{l\text{-}s} = w_s \langle\Sigma^{-1}_\text{crit}\rangle^2(z).
\end{equation}
In equation \eqref{eq:wls}, $w_s$ is the inverse variance weight for the galaxy shape in
our source catalogue, $z$ is the redshift of the lens, and
\begin{equation}\label{eq:sigma_crit_phot_z}
    \langle\Sigma^{-1}_\text{crit}\rangle^2(z) =
    \left[\int \Sigma_\text{crit}^{-1}(z, z_s)\, n(z_s) \, dz_s\right]^2
\end{equation}
is the photo-\z corrected (squared inverse) critical surface density \citep{Sheldon2004}.
Here, $n(z_s)$ is the distribution from Section \ref{sec:sources} and
\begin{equation}
  \Sigma_\text{crit}^{-1}(z, z_s)
  = \frac{4\pi G}{c^2} \frac{D_\mathrm{A}(z) D_\mathrm{A}(z, z_s)}{D_\mathrm{A}(z_s)}
\end{equation}
is the standard (inverse) critical surface density that is dependent only on angular
diameter distances---that is, only dependent on the geometry---between observer and lens,
$D_\mathrm{A}(z)$, observer and source, $D_\mathrm{A}(z_s)$, and between lens and source,
$D_\mathrm{A}(z, z_s)$.

We improve the \dsigma measurement of equation \eqref{eq:shear} by including the boost
factor and random subtraction, so the full estimator of excess surface density becomes
\citep{Lange2022}:
\begin{equation}
  \dsigma(R) = b(R)\dsigma_\text{true lens}(R) - \dsigma_\text{random lens}(R).
\end{equation}
In the formula above, $\dsigma_\text{true/random lens}$ is the lensing signal calculated
around the true/random lenses, and
\begin{equation}
  b(R) = \frac{%
    \sum_{l\text{-}s}w_{\text{sys},l}w_{l\text{-}s}}{%
    \sum_{r\text{-}s}w_{\text{sys},r}w_{r\text{-}s}%
  }
\end{equation}
is the radially-dependent boost factor. $\sum_{r\text{-}s}$,  $w_{\text{sys},r}$, and
$w_{r\text{-}s}$ mean the same as $\sum_{l\text{-}s}$, $w_{\text{sys},l}$, and
$w_{l\text{-}s}$, respectively, but they refer to the random lenses as opposed to the
(true) lenses in our sample. The $w_{\text{sys},r}$ weights are all unity.

Lastly, the uncertainties in the measured \dsigma signal come from jackknifing
$N_\text{JK}=100$ different patches of the sky, which accounts for physically correlated
noise (e.g., due to foreground large-scale structures). Any uncertainties that arise from
our various correction factors are propagated through to the final \dsigma uncertainty
since the jackknife error estimate is determined after all corrections are applied.

These \dsigma measurements are performed in $N_\text{b}=14$ linearly-spaced radial bins of
equal width spanning $R=0.1\Mpc$ to $R=1.5\Mpc$. The upper limit of $R=1.5\Mpc$ should be
sufficiently large to enclose the DM haloes of our galaxies, which have stellar masses of
$\sim$$10^{11}\Msun$. The lower limit of $R=0.1\Mpc$ is to avoid biases that arise when
shape measurements are made too close to the centre of lenses.

\subsection{Fitting Halo Profiles}\label{sec:fitting}

Finally, we fit Navarro-Frenk-White (NFW) profiles to the \dsigma measurements to
determine whether the merger process produces measurable differences in the DM haloes of
galaxies. In our analysis, we adopt a virial overdensity threshold that varies with
redshift as given by \citet{Bryan1998} and implemented in the \textsc{colossus} Python
package \citep{Diemer2018}. For our lenses, which have an average redshift of
$\lzr\approx0.15$, this virial overdensity corresponds to $\sim$114$\times$ the critical
density of the Universe. This overdensity threshold defines the edge of our halo radius
\rvir and therefore the enclosed halo mass \mhalo and concentration \c.

Our full model consists of two components:~an NFW halo and a point-like mass of stars at
the centre of the galaxy. The stellar component has a mass equal to the weighted
arithmetic mean of the stellar masses in the given lens sample, with weights given by
equation \eqref{eq:weights} for controls and unity weights for post-mergers. The DM
component has two free parameters that we fit to the data:~the halo mass \mhalo and
concentration \c. The small ($<\,$2\%) uncertainties in the stellar masses estimated from
10\,000 bootstrapping iterations justify our choice to keep \Mstar as a fixed parameter
(see Table \ref{tab:hyp_test_fit} in Section \ref{sec:results}), allowing for smaller
uncertainties in the free parameters. Changing the stellar masses by $\pm20\%$ to account
for systematic uncertainties  in the stellar population synthesis models
\citep[e.g.,][]{Roediger2015, Dogruel2023} only affects our nominal halo masses and
concentrations by $\lesssim$\,5\%---well within the \mhalo and \c uncertainties given in
Section \ref{sec:results}'s Table \ref{tab:hyp_test_fit}.

We find the best-fitting \mhalo and \c to our \dsigma measurements using a Markov chain
Monte Carlo routine to minimize the following $\chi^2$:
\begin{equation}\label{eq:chisq}
  \chi^2 = \left(\boldsymbol{\dsigma} - \boldsymbol{\widehat{\dsigma}}\right)^\mathrm{T}
  \mathbf{C}^{-1}\!
  \left(\boldsymbol{\dsigma} - \boldsymbol{\widehat{\dsigma}}\right),
\end{equation}
where $\boldsymbol{\dsigma}$ is a vector containing our measurements from Section
\ref{sec:measurements}, $\boldsymbol{\widehat{\dsigma}}$ is a vector containing the
predicted \dsigma amplitudes using the current parameters, and $\mathbf{C}^{-1}$ is the
unbiased inverse covariance matrix of our \dsigma values accounting for the Hartlap factor
\citep{Hartlap2007}. We adopt uninformed, uniform priors for the halo mass and
concentration, where the log-prior probabilities are zero for $\mhalo\geq10^9\Msun$ and $0
\leq c \leq 10$, and negative infinity otherwise.

Note that in our analysis, we only model single NFW haloes without any additional
contributions from other galaxies or from the environment (such as the underlying dark
matter halo of a galaxy cluster). As discussed in Section \ref{sec:env_control}, although
the SDSS DR7 spectroscopic catalogue is extensive, our environment control procedure is
imperfect because fibre collisions may cause undercounting of the number of neighbouring
galaxies around a lens. Moreover, we estimate three-dimensional distances from redshift by
assuming galaxies move only according to the Hubble flow. Galaxies in dense clusters with
relatively large peculiar motions may therefore be mistaken to be far apart simply because
we assume differences in redshift are entirely due to cosmological expansion. With these
considerations, we only fit to data points below $R\leq1.0\Mpc$, since beyond this
distance, the 2-halo term becomes important \citep[figure 11]{Cooray2002}.

\section{Results and Discussion}\label{sec:results}

\begin{figure}
  \centering
  \centering
  \includegraphics[width=\columnwidth]{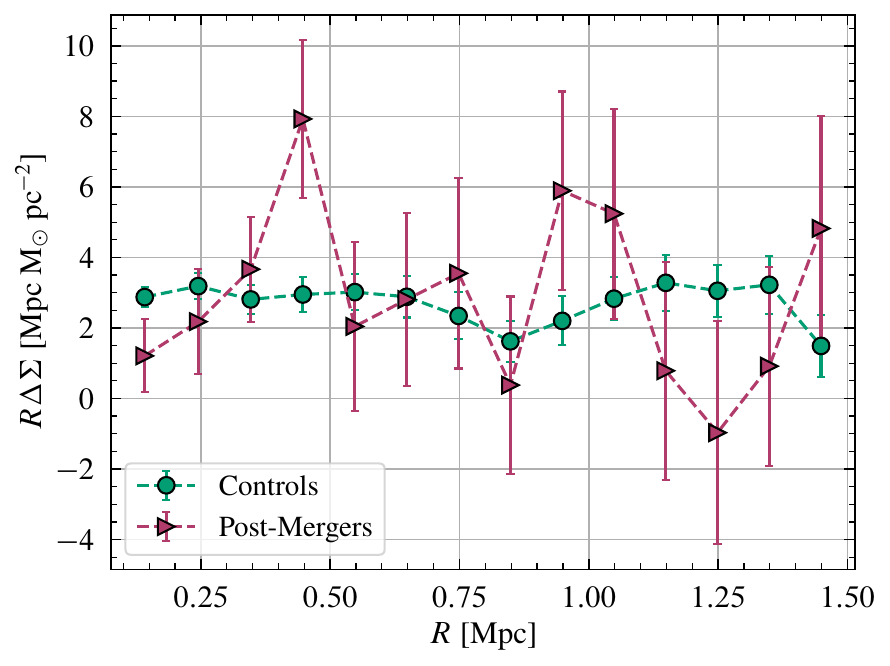}
  \caption{The \rdsigma signal of our lenses (post-mergers and non-merging controls).
  $R$ is the distance from the centre of the lens and \dsigma is the excess surface
  density. The lensing amplitudes include boost correction and random subtraction.}
  \label{fig:dsigma_results_envControl}
\end{figure}

\begin{figure*}
  \centering
  \hfill
  \begin{subfigure}[t]{0.35\textwidth}
  \vspace*{0pt}
  \centering
  \includegraphics[width=\textwidth]{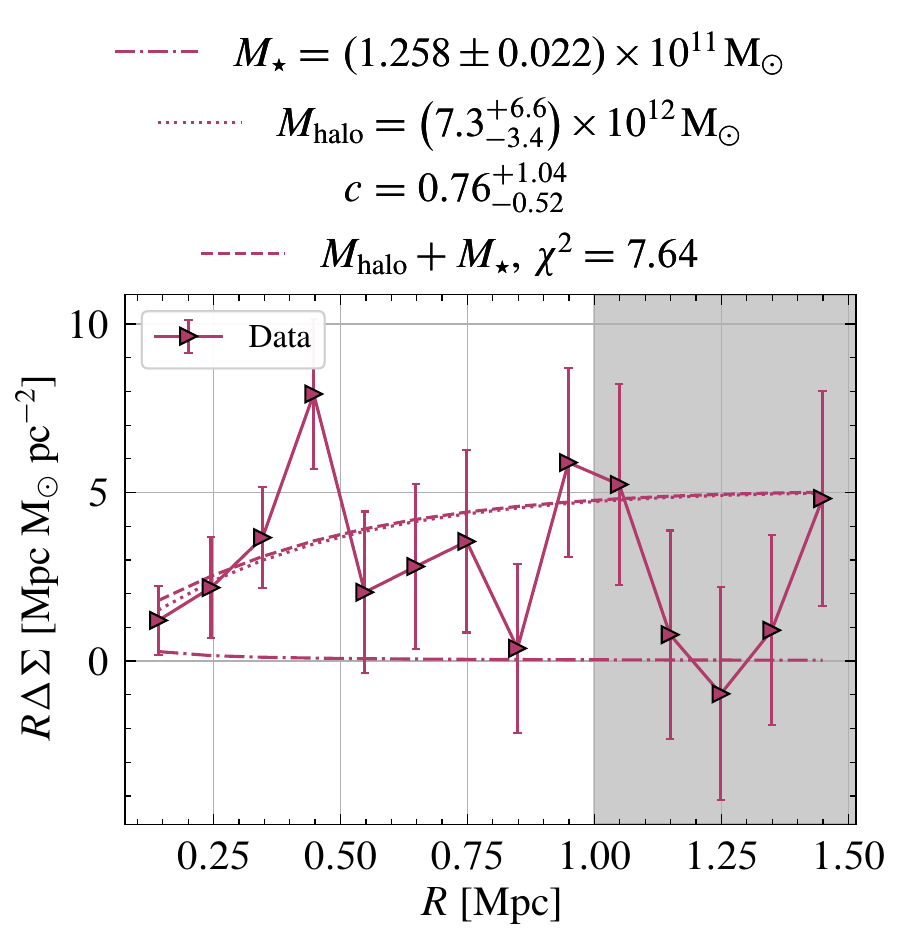}
  \includegraphics[width=\textwidth]{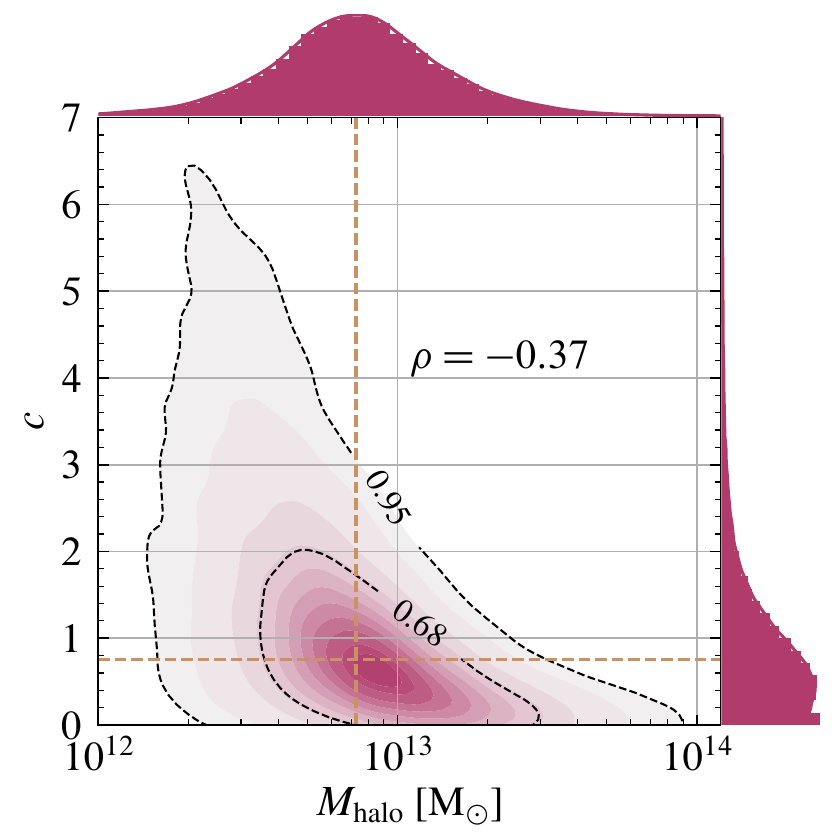}
  \caption{Post-mergers}
  \label{subfig:post_merger_fits}
  \vspace*{0pt}
  \end{subfigure}
  \hfill
  \begin{subfigure}[t]{0.35\textwidth}
  \vspace*{0pt}
  \centering
  \includegraphics[width=\textwidth]{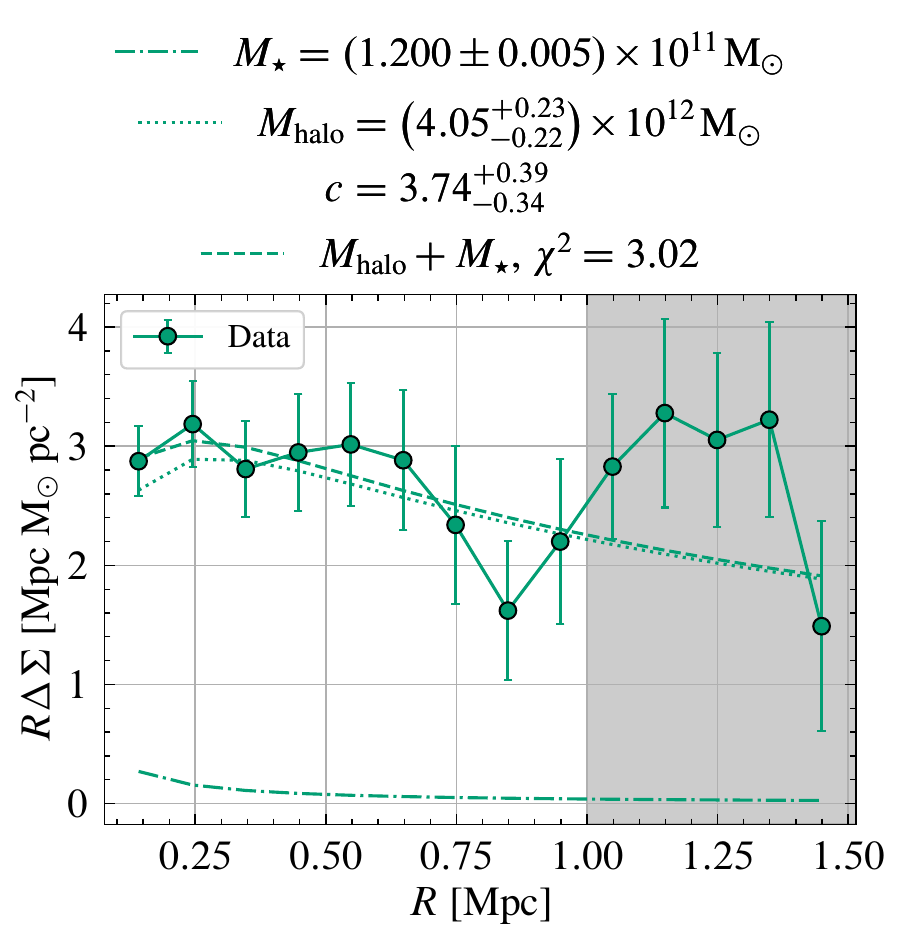}
  \includegraphics[width=\textwidth]{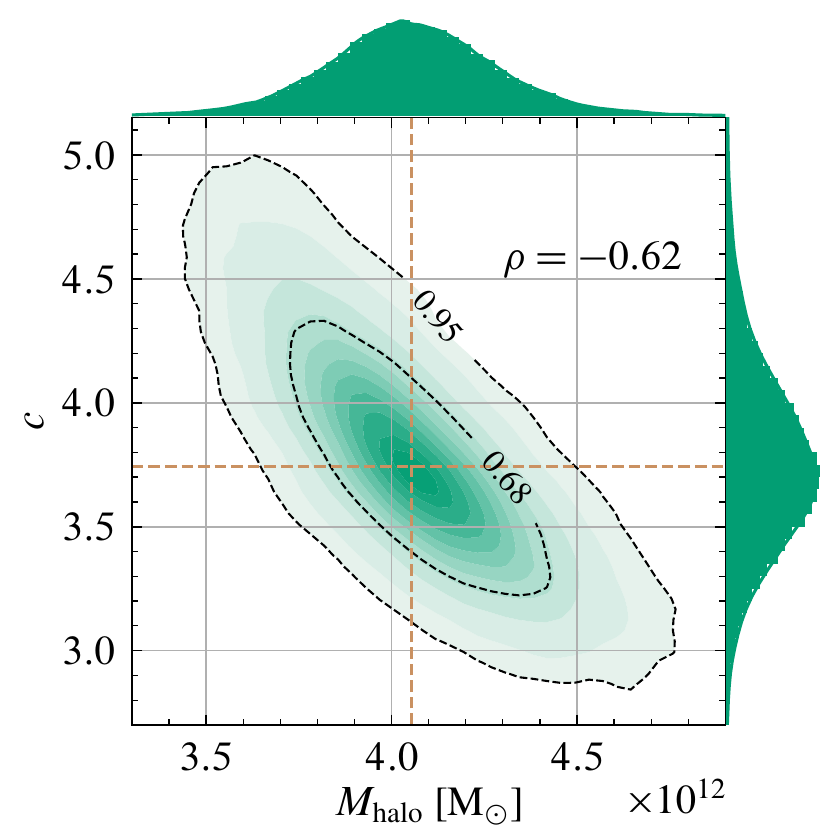}
  \caption{Non-merging controls}
  \label{subfig:control_fits}
  \vspace*{0pt}
  \end{subfigure}
  \hfill
  \vspace*{6pt}
  \caption{NFW halo fits to the lensing signal and correlations in the fitted parameters.
  The lower panels show the Pearson product-moment correlation coefficient $\rho$ between
  the concentration $c$ and the halo mass \mhalo, with the 68 and 95 percentile contours
  indicated. The nominal (i.e., median) values of these parameters are marked with dashed
  gold lines. Note that the fit and all its derived parameters (including the
  uncertainties, $\chi^2$, and $\rho$) only consider the data points not in the shaded
  region, as discussed in Section \ref{sec:fitting}. The stellar masses quoted in the top
  plots do not include systematic uncertainties in the stellar population modelling, but
  the impact of these systematic effects is minimal on our derived halo masses and
  concentrations (see Section \ref{sec:fitting}).}
  \label{fig:fits}
\end{figure*}

We now present the results of our analysis detailed in Section \ref{sec:methods}. Figure
\ref{fig:dsigma_results_envControl} shows the \dsigma measurements using our lenses from
Figure \ref{fig:all_distr_matched}. We remark that the largest boost correction increases
\dsigma by $\sim$10\% in the smallest radial bin, gradually decreasing to corrections of
less than 3.5\% for $R\in[1.4, 1.5] \Mpc$, in agreement with previous work
\citep[e.g.,][figure 3, first column]{Prat2022}.

Additionally, Figure \ref{fig:dsigma_results_envControl} suggests that the lensing signal
of our post-merger catalogue is not significantly different from that of the controls.
Indeed, we can perform a chi-square test to examine the null hypothesis that the \dsigma
profile from a given sample is \emph{not} significantly different from that of another
sample. For our $\chi^2_{1,2}=14.50$, we obtain $p=0.41$, i.e., our post-mergers and
controls do not produce significantly different lensing amplitudes. Note that comparing
only the 9 data points where $R \leq 1.0\Mpc$ does not appreciably change the $p$-value
(which becomes $p=0.34$).

Next, we fit an NFW halo to each of our datasets following the procedure given in Section
\ref{sec:fitting}. Figure \ref{fig:fits} shows the best-fitting models as well as
constraints on---and correlations between---the fitted parameters. We check that the
fitted models are actually \emph{good} fits by performing a chi-square test and find that
the data and model are mutually consistent. Our best-fitting parameters are summarized in
Table \ref{tab:hyp_test_fit}.

\begin{table*}
    \renewcommand{\arraystretch}{1.2}
    \centering
    \caption{Summary table including number of data points $N$ in each sample, mean
    redshift, parameters in our model, SHMR, and goodness-of-fit hypothesis testing
    results for the null hypothesis that the data and model are consistent, with 7 DoF.
    The stellar masses do not include systematic uncertainties in the stellar population
    modelling, but the impact of these systematic effects is minimal on our derived halo
    masses and concentrations (see Section \ref{sec:fitting}).}
    \label{tab:hyp_test_fit}
    \vspace{-0.5\baselineskip}
    \begin{tabular}{cccccccccc}
      \toprule
      Sample & $N$ & \lzr & \Mstar ($\times \text{10}^\text{11} \Msun$) & \mhalo ($\times
      \text{10}^\text{12} \Msun$) & \c & $\Mstar/\mhalo$ (\%) & $\chi^2$ & $p$-value \\
      \midrule
      Post-Mergers  & 1\,623     & $\text{0.1525}\pm\text{0.0012}$ &
      $\text{1.258}\pm\text{0.022}$ & $\text{7.3}\errs{6.6}{3.4}$   &
      $\text{0.76}\errs{1.04}{0.52}$ & $\text{1.7}\errs{1.6}{0.8}$ & 7.64 & 0.37 \\
      Controls      & 30\,048.76 & $\text{0.1508}\pm\text{0.0003}$ &
      $\text{1.200}\pm\text{0.005}$ & $\text{4.05}\errs{0.23}{0.22}$ &
      $\text{3.74}\errs{0.39}{0.34}$ & $\text{2.96}\errs{0.17}{0.16}$ & 3.02 & 0.88 \\
      \bottomrule
    \end{tabular}
\end{table*}

Figure \ref{fig:fits} reveals that \mhalo is moderately negatively correlated with \c in
the fits to both datasets. This is because decreasing the concentration pushes the mass
out to larger radial scales; to keep the lensing signal the same (i.e., at small
$R\leq1.0\Mpc$), the halo mass must therefore increase to compensate for the lower
concentration. We now examine each of these parameters in more detail.

At the 95\% confidence level, the best-fitting halo concentrations to our
post-mergers ($\c=0.76\errs{1.04}{0.52}$) and controls ($\c=3.74\errs{0.39}{0.34}$) can
range between $[0.03, 5.30]$ and $[2.99, 4.68]$, respectively. Although previous work has
shown that concentrations in post-merger systems may be lower than those of non-mergers by
up to a factor of $\approx\,$2 \citep[depending on how long it has been since the merger
event, see][figure 4]{Wang2020}, our data do not provide statistically significant
constraints on how mergers can affect the DM structure of galaxies.

Moreover, these best-fitting concentrations are somewhat smaller than those predicted from
simulations. For instance, our concentrations are not consistent with the
\citet{Duffy2008} concentration-mass relation (hereafter Duffy08), which predicts
concentrations of 6.58 and 6.91 for halo masses equal to the best-fitting \mhalo of the
post-mergers and controls, respectively. These correspond to discrepancies of
$\approx\,$2.1$\sigma$ for the post-mergers and $\approx\,$3.2$\sigma$ for the controls.
One possible explanation is that Duffy08 is calibrated using dark matter only (DMO)
simulations, but baryonic feedback in galaxies pushes mass out to larger radii, thereby
decreasing concentrations relative to Duffy08 \citep{Duffy2010, Shao2023}.

When feedback is included, \citet{Duffy2010} show that the concentration of a galaxy's DM
component decreases by about 15\% compared to an equivalent DMO simulation with no
feedback, so we might expect $c\approx6.91\times0.85\approx5.87$ for the DM component.
There is more to the story, however, as our lensing measurements are sensitive to not only
the DM but also the gas in our samples. The mass in $\mhalo\sim4\times10^{12}\Msun$
galaxies consists of $\approx\,$85\% DM, $\approx\,$2\% stars, and $\approx\,$13\% gas. It
is this gas that is pushed out to large radii by feedback, so the concentration of the
combined DM and gas will be even lower than DM alone.

Alternatively, the low concentrations may be partly due to the presence of satellite
galaxies in our sample. If some objects in our dataset are indeed satellites (i.e., with a
nearby, more massive parent halo that we have missed), then the measured lensing profile
will be affected by the off-centre host haloes, reducing the apparent concentration
\citep[e.g.,][]{Mandelbaum2005, Yang2006, Du2014}. Our isolation criteria (Section
\ref{sec:env_control}) limits this possibility, but because of fibre collisions, we cannot
completely rule it out.\footnote{Another possibility is that there may be satellite
galaxies around the haloes in our sample. In this case, however, as long as we have
identified the main halo ($\Mstar \sim 10^{11}\Msun$) correctly, the contribution of
satellite galaxies to the main halo's lensing signal is minimal since only $\sim$10\% of
the total mass is in satellites around such massive galaxies.}. Moreover, given the scales
at which our measurements are made---$R\leq1\Mpc=0.7\,h^{-1}\Mpc$---we still expect the
lensing signal to be dominated by central haloes \citep[figure 1]{Mandelbaum2005} if the
satellite fraction is $\lesssim$\,20\%. However, the full impact of possible satellite
galaxy contamination in our samples is difficult to determine without simulations and mock
observations that incorporate fibre collisions; we therefore leave a more detailed
investigation of these effects to future work.

The low concentrations may also be due to halo assembly bias. It is well established that
galaxies in lower density environments tend to form later and have lower concentrations
than their counterparts in more overdense regions \citep[e.g.,][]{Wechsler2002,
Maccio2007}. For our halo masses of $\mhalo\sim4\times10^{12}\Msun \sim
10^{12.4}\,h^{-1}\Msun$, however, \citet[figure 10]{Maccio2007} show that this dependence
is weak or even slightly reversed (i.e., slightly higher concentrations in underdense
regions). A more careful analysis of the impact of halo assembly bias on weak lensing
measurements is beyond the scope of this paper and is deferred to later work.

Table \ref{tab:hyp_test_fit} also contains the SHMRs of our various datasets. By comparing
the SHMRs of our post-mergers to our controls, we can determine if sustained, elevated SF
occurs during the merger process that is sufficient to produce a detectable difference in
their SHMRs. While the SHMRs themselves may be subject to systematic errors, the
\emph{ratio} of the SHMRs of these matched samples should be robust because systematics
will affect the halo masses in the same way. Our observed ratio of the post-merger SHMR to
control SHMR is $\ratio=0.58\errs{0.53}{0.28}$ (cf.~Table \ref{tab:hyp_test_fit}). Using
the models from \citet{Hudson2015} and the average stellar mass of the post-mergers in
Table \ref{tab:hyp_test_fit}, our results rule out extreme ($\gtrsim\,$60\%) bursts of SF
at the 95\% confidence level. This perhaps indicates that merger events do not, in
general, yield galaxies with extreme SHMRs relative to non-mergers. Indeed,
\citet{Ferreira2024_oct} find that galaxies with $\Mstar\sim10^{11}\Msun$ only increase
their final stellar masses by $\approx\,$10\% after a merger \citep[see
also][]{Reeves2024}. Such a burst would predict $\ratio \sim 1.4$, consistent with our
result.

To detect a starburst of similar ($\approx\,$10\%) magnitude using weak lensing, we would
require a sample of $\sim$15\,000 post-mergers to reduce our observed error in $\ratio$ to
$\lesssim\,$0.2. This is about 10 times larger than the current sample. Assuming 5\% of
observed galaxies are merger products \citep{Casteels2014,Ferreira2024}, the full dataset
would therefore need to have $\sim$300\,000 galaxies with individual redshift
measurements.

A post-merger sample of this size is most likely to come from either lenses that only have
photometric, not spectroscopic, redshifts or from surveys by \textit{Euclid} or the Dark
Energy Spectroscopic Instrument (DESI). In addition to a larger post-merger sample, DESI
should also be able to obtain redshifts and other measurements for galaxies that are
within 55\arcsec of each other since its survey strategy observes the same patches of sky
multiple times. This is in contrast to SDSS, where most areas in the sky are only covered
by one pointing so galaxies that are too close together only have a single redshift due to
fibre collisions, as in our current catalogue (also see the discussion in Section
\ref{sec:env_control}). The DESI data would therefore make our environment control more
robust, since neighbouring galaxies that were previously missed are now accounted for.
Finally, DESI surveys are also deeper than SDSS, which should provide lenses at a more
optimal redshift for UNIONS source galaxies.

\section{Conclusions}\label{sec:conclusions}

In this paper, we use galaxy-galaxy lensing to investigate how mergers change the DM
haloes and stellar content of galaxies. Our lens galaxies come from a catalogue of
candidate mergers taken from UNIONS and SDSS DR7, while our source galaxies come from the
UNIONS ShapePipe v1.3 catalogue. To ensure differences in the lensing measurements are due
to the merger process, we weight the post-mergers and non-merging control galaxies to have
the same distributions of stellar mass \Mstar, redshift \z, and geometric mean distance to
their three nearest neighbours \ravg. Additionally, we only use galaxies that live in
low-density environments to reduce the effect of the 2-halo term. Our final sample
consists of 1\,623 post-mergers and $\approx\,$30\,000 controls.

Using these data, we measure their \dsigma amplitudes and find that our post-mergers and
non-merging controls produce similar lensing signals. We adopt a two-component model to
fit to the lensing measurements below $R\leq1.0\Mpc$ consisting of a fixed point-like
stellar mass contribution and an NFW halo with halo mass and concentration as free
parameters. With our current data, we do not find statistically significant differences in
the DM haloes of galaxies, nor in their stellar fractions, as they progress along the
merger sequence. NFW haloes fitted to the two sets of lensing measurements both have
$\mhalo\sim4\times10^{12}\Msun$ and show moderately negative correlation between halo mass
and concentration. The best-fitting post-merger concentration of
$\c=0.76\errs{1.04}{0.52}$ is not significantly different from (i.e., within 2$\sigma$ of)
the best-fitting $\c=3.74\errs{0.39}{0.34}$ of the controls. These concentrations are
smaller than those predicted from simulations, but this may be due to differences between
feedback in nature and feedback in simulations, possible satellite galaxy contamination,
halo assembly bias, and/or limitations of our two-component model. We find that the
post-mergers and controls have SHMRs of $1.7\errs{1.6}{0.8}$\% and
$2.96\errs{0.17}{0.16}$\%, respectively. Our SHMRs place an upper limit ($\lesssim$\,60\%)
on the fraction of stellar mass produced in the galaxy merger process; this result is in
agreement with previous studies that show merger events do not greatly affect post-merger
stellar masses.

We have shown that weak lensing is a viable method to study the properties of galaxy
mergers. With a more complete dataset of merger products, or of weak lensing source
galaxies, such as from DESI, Rubin/LSST, \textit{Euclid}, and \textit{Roman}, we may yet
uncover more of the complex effects that galaxy mergers have on their progenitors using
galaxy-galaxy lensing.


\section*{Data Availability}

The merger catalogue from \citet{Ferreira2024} is publicly available on
GitHub\footnote{\url{https://github.com/astroferreira/MUMMI_UNIONS}} with redshifts from
SDSS DR7 and stellar masses from MPA-JHU (as discussed in the text).

A subset of the raw data underlying the source catalogue used in this article is publicly
available via the Canadian Astronomy Data
Centre\footnote{\url{http://www.cadc-ccda.hia-iha.nrc-cnrc.gc.ca/en/megapipe/}}. The
remaining raw data and all processed data are available to members of the Canadian and
French communities via reasonable requests to the principal investigators of the
Canada-France Imaging Survey, Alan McConnachie and Jean-Charles Cuillandre. All data will
be publicly available to the international community at the end of the proprietary period.


\begin{acknowledgments}

We are honoured and grateful for the opportunity of observing the Universe from Maunakea
and Haleakala, which both have cultural, historical and natural significance in Hawaii.
This work is based on data obtained as part of the Canada-France Imaging Survey, using
observations obtained with MegaPrime/MegaCam, a joint project of the Canada-France-Hawaii
Telescope (CFHT) and CEA Saclay, on the CFHT, which is operated by the National Research
Council of Canada, the Institut National des Science de l'Univers of the Centre National
de la Recherche Scientifique of France, and the University of Hawaii. This research is
based in part on data collected at Subaru Telescope, which is operated by the National
Astronomical Observatory of Japan. Pan-STARRS is a project of the Institute for Astronomy
of the University of Hawaii, and is supported by the NASA SSO Near Earth Observation
Program under grants 80NSSC18K0971, NNX14AM74G, NNX12AR65G, NNX13AQ47G, NNX08AR22G,
80NSSC21K1572, and by the State of Hawaii.

This research used the facilities of the Canadian Astronomy Data Centre operated by the
National Research Council of Canada with the support of the Canadian Space Agency.
Additionally, we are grateful for the computing resources of the Digital Research Alliance
of Canada\footnote{\url{https://ccdb.alliancecan.ca/}} and the CANFAR (Canadian Advanced
Network for Astronomical Research) Science
Portal\footnote{\url{https://www.canfar.net/science-portal/}} for enabling the analysis in
this paper.

MJH acknowledges support from NSERC through a Discovery Grant. HH is supported by a DFG
Heisenberg grant (Hi 1495/5-1), the DFG Collaborative Research Center SFB1491, an ERC
Consolidator Grant (No. 770935), and the DLR project 50QE2305.

We also thank the anonymous referee for their insightful feedback, particularly regarding
factors that may contribute to the measured low concentrations.

\end{acknowledgments}


\begin{contribution}

IC performed the analysis and wrote the manuscript, with guidance from JE-P and MJH. MJH
additionally calculated the theoretical SHMRs to compare with our observed ratios. All
other coauthors contributed to the various data products used in this paper and/or
provided significant feedback to improve this work.

\end{contribution}


\bibliography{citations}{}
\bibliographystyle{aasjournalv7}


\end{document}